\documentclass[twocolumn,showpacs,preprintnumbers,amsmath,amssymb,superscriptaddress,nofootinbib]{revtex4}
\begin{document}
\topmargin -2cm
\title{Viscous dark energy and phantom evolution}
\author{Mauricio Cataldo}
\altaffiliation{mcataldo@ubiobio.cl}%0915
\affiliation{Departamento de F\'\i sica, Facultad de Ciencias,
Universidad del B\'\i o--B\'\i o, Avenida Collao 1202, Casilla
5-C, Concepci\'on, Chile.\\}
\author{Norman Cruz}
\altaffiliation{ncruz@lauca.usach.cl} \affiliation{Departamento de
F\'\i sica, Facultad de Ciencia, Universidad de Santiago, Casilla
307, Santiago, Chile.\\}
\author{Samuel Lepe}
\altaffiliation{slepe@ucv.cl} \affiliation{Instituto de F\'\i
sica, Facultad de Ciencias B\'asicas y Matem\'aticas, Pontificia
Universidad Cat\'olica de Valpara\'\i so, Avenida Brasil 2950,
Valpara\'\i so, Chile.}
%\date{\today, \bf { file: articles02-fridman.prlnfrwfin.tex}}
\begin{abstract}
In order to study if the bulk viscosity may induce a big rip
singularity on the flat FRW cosmologies, we investigate
dissipative processes in the universe within the framework of the
standard Eckart theory of relativistic irreversible
thermodynamics, and in the full causal Israel–Stewart-Hiscock
theory. We have found cosmological solutions which exhibit, under
certain constraints, a big rip singularity. We show that the
negative pressure generated by the bulk viscosity cannot avoid
that the dark energy of the universe to be phantom energy.

\vspace{0.5cm}
\end{abstract}
\maketitle
\section{Introduction}

The existence of a exotic cosmic fluid with negative pressure,
which constitutes about the $70$ per cent of the total energy of
the universe, has been perhaps the most surprising discovery made
in cosmology. This dark energy is supported by the astrophysical
data obtained from Wilkinson Microwave Anisotropy Probe (WMAP)
(Map) and high redshift surveys of supernovae.

The dark energy is considered a fluid characterized by a negative
pressure and usually represented by the equation of state $w= p/
\rho$, where $w$ lies very close to $-1$, most probably being
below $-1$. Dark energy with $w<-1$, the phantom component of the
universe, leads to uncommon cosmological scenarios as it was
pointed out in~\cite{Cadwell}. First of all, there is a violation
of the dominant energy condition (DEC), since $\rho+ p <0$. The
energy density grows up to infinity in a finite time, which leads
to a big rip, characterized by a scale factor blowing up in this
finite time. These sudden future singularities are, nevertheless,
not necessarily produced by a fluids violating DEC.
Barrow~\cite{Barrow 1} has shown, with explicit examples, that
exist solutions which develop a big rip singularity at a finite
time even if the matter fields satisfy the strong-energy
conditions $\rho>0$ and $\rho +3p >0$.  A generalization of
Barrow's model has been realized in~\cite{Nojiri}, giving its
Lagrangian description in terms of scalar tensor theory.  It was
also proved by Chimento et al~\cite{Chimento} that exist a duality
between phantom and  flat Friedmann-Robertson-Walker (FRW)
cosmologies with nonexotic fluids. This duality is a
form-invariance transformation which can be used for constructing
phantom cosmologies from standard scalar field universes.
Cosmological solutions for phantom matter which violates the weak
energy condition were found in~\cite{Dabrowski}.

The role of the dissipative processes in the evolution of the
early universe also has been extensively studied. In the case of
isotropic and homogeneous cosmologies, any dissipation process in
a FRW cosmology is scalar, and therefore may be modelled as a bulk
viscosity within a thermodynamical approach.

A well known result of the FRW cosmological solutions,
corresponding to universes filled with perfect fluid and bulk
viscous stresses, is the possibility of violating DEC~\cite{Barrow
2}. The bulk viscosity introduces dissipation by only redefining
the effective pressure, $P_{_{eff}}$, according to
\begin{eqnarray}\label{effecti}
P_{_{eff}}= p+\Pi= p-3 \xi H,
\end{eqnarray}
where $\Pi$ is the bulk viscous pressure, $\xi$ is the bulk
viscosity coefficient and $H$ is the Hubble parameter. Since the
equation of energy balance is
\begin{eqnarray}\label{ConsEq}
\dot{\rho}+ 3 H (\rho+p+\pi)=0.
\end{eqnarray}
the violation of DEC, i.e., $\rho+p+\Pi<0$ implies an increasing
energy density of the fluid that fills the universe, for a
positive bulk viscosity coefficient. The condition $\xi >0$
guaranties a positive entropy production and, in consequence, no
violation of the second law of the thermodynamics~\cite{Pavon}.

In the present paper we show that the above results are
straightforward to obtain from the exact cosmological solutions
already found by Barrow in~\cite{Barrow 2}. These solutions were
obtained using non causal thermodynamics. Nevertheless, we
consider a more physical approach like the full
Israel-Stewart-Hiscock causal thermodynamics, showing that it is
also possible to obtain big rip type solutions.

The organization of the paper is as follows: in Section II we
present the field equations for a flat FRW universe filled with a
bulk viscous fluid within the framework of the Eckart theory. We
indicate that under a constraint for the parameters of the fluid,
one of the Barrow's solutions presents a future singularity in a
finite time. In Section III we obtain big rip solutions using the
approach of the full Israel-Stewart-Hiscock causal thermodynamics.
In Section IV we discuss our results in relation to the nature of
the dark energy of the universe.

\section{Eckart theory}
The FRW metric for an homogeneous and isotropic flat universe is
given by
\begin{eqnarray}\label{ndim}
ds^2=-dt^2+a(t)^2\left(dr^2+r^2 (d\theta^2 + sin^2 \theta d
\phi^2) \right),
\end{eqnarray}
where $a(t)$ is the scale factor and $t$ represents the cosmic
time. In the following we use the units $8\pi G=1$. In the first
order thermodynamic theory of Eckart~\cite{Eckart} the field
equations in the presence of bulk viscous stresses are
\begin{eqnarray}\label{tt}
\left(\frac{\dot{a}}{a}\right)^2=H^2= \frac{\rho}{3},
\end{eqnarray}
\begin{eqnarray}\label{rr}
\frac{\ddot{a}}{a}=\dot{H}+H^2=-\frac{1}{6} \left(\rho + 3
P_{_{eff}} \right),
\end{eqnarray}
with
\begin{eqnarray}\label{Peff}
P_{_{eff}}=p+\Pi,
\end{eqnarray}
and
\begin{eqnarray}\label{xi}
\Pi=-3 H \xi.
\end{eqnarray}

The conservation equation is
\begin{eqnarray}\label{ConsEq}
\dot{\rho}+ 3 H (\rho+p+\Pi)=0.
\end{eqnarray}
Assuming that the dark component obey the state equation
\begin{eqnarray}\label{gamalaw}
p=(\gamma-1) \rho,
\end{eqnarray}
where $0\leq  \gamma \leq 2$, we can obtain from
equations~(\ref{tt}) to~(\ref{gamalaw}) a single evolution
equation for $H$:
\begin{eqnarray}\label{Hpunto}
2\dot{H}+ 3 \gamma H^{2}=3\xi H.
\end{eqnarray}

This equation may be integrated directly as a function of the bulk
viscosity. For $\gamma \neq 0$ the solution has the form
\begin{eqnarray}
H(t)=\frac{e^{\frac{3}{2}\int \xi(t) dt}}{C+\frac{3}{2} \gamma
\int e^{\frac{3}{2}\int \xi(t) dt} \, dt}
\end{eqnarray}
where $C$ is an integration constant. From this equation we find
the following expression for the scale factor
\begin{eqnarray}\label{scalefactor1}
a(t)= D \left(C+\frac{3}{2} \, \gamma \int e^{\frac{3}{2} \, \int
\xi(t) dt} \, dt \right)^{2/(3 \gamma)},
\end{eqnarray}
where $D$ is a new integration constant. Thus for a given $\xi(t)$
we have the expressions for $a(t)$, $\rho(t)$ and $p(t)$.

For the case $\gamma=0$ we have from Eq.~(\ref{Hpunto})
\begin{eqnarray}\label{bulkv}
\xi=\frac{2}{3} \, \frac{\dot{H}}{H},
\end{eqnarray}
and substituting this expression into Eq.~(\ref{ConsEq}) we have
\begin{eqnarray}\label{rhodot}
 \dot{\rho}= 6 \dot{H} H,
\end{eqnarray}
from which we conclude that $\rho=3H^2+const$. Comparing this
expression with~(\ref{tt}) we have that the integration constant
is zero.

Thus we have that for the state equation $p=-\rho$, i.e. for
$\gamma=0$, the scale factor is not defined by the field
equations. So for a given $a(t)$ we can write $H$ and then obtain
the expressions for the energy density from Eq.~(\ref{tt}) and the
bulk viscosity from Eq.~(\ref{bulkv}). Clearly if $\xi=0$ the well
known de~Sitter scale factor $a(t)=e^{H_{_{0}} t }$ is obtained,
where $p=-\rho$ and both are constants.

Notice that the solution of the field equations may be written
through $\xi(t)$ or $a(t)$ because there are three independent
equations for the four unknown functions $a(t)$, $\rho(t)$,
$\xi(t)$ and $p(t)$.

Now we are interested in the possibility that there are
cosmological models with viscous matter which present in its
development a big rip singularity.

\subsection{The case for $\gamma \neq 0$}

Firstly, let us consider the case $\gamma \neq 0$. If the viscous
fluid satisfies DEC, then the condition $0 \leq \gamma \leq 2$
must be satisfied. Thus for $\gamma < 0$ we have a phantom
cosmology. Now from the thermodynamics we know that $\xi > 0$, and
if $\gamma < 0$ the Eq.~(\ref{scalefactor1}) implies that we can
have a big rip singularity at a finite value of cosmic time.

Let us consider some examples to see this more clearly. From
Eq.~(\ref{scalefactor1}) the well known standard case for a
perfect fluid, i.e. $\xi=0$, takes the form $a(t)=D(C+(3/2) \gamma
t)^{2/(3 \gamma)}$. This scale factor may be rewritten as
\begin{eqnarray}
a(t)=a_{_{0}} \left(1+\frac{3}{2} H_{_0}\gamma t
\right)^{2/(3\gamma)},
\end{eqnarray}
and the energy density is given by
\begin{eqnarray}
\rho=\frac {\rho_{_{0}}}{ \left( 1+\frac{3}{2} H_{_0} \gamma t
\right) ^{2}},
\end{eqnarray}
where $\rho_{_{0}}=3 H^2_{_0}$, in order to have
$H(t_{_0}=0)=H_{_0}>0$. If $\gamma <0$ we have a big rip
singularity at a finite value of cosmic time $t_{_{br}}=-2/( 3
H_{_0} \gamma) > t_{_{0}}=0$.

In the special case of $\xi(t)=\xi_{_{0}}=const$ we have from
Eq.~(\ref{scalefactor1}) for the scale factor
$a(t)=D(C+(\gamma/\xi_{_{0}}) \, e^{(3/2)\xi_{_{0}}t}
)^{2/(3\gamma)}$. We can rewrite it into the form
\begin{eqnarray}
a(t)=a_{_{0}} \left( 1+\frac{H_{_0}}{\xi_{_0}}\, \gamma \left(
e^{\, 3 \, \xi_{_{0}}\, t/2}-1 \right)\right)^{2/(3 \gamma)},
\end{eqnarray}
from which we obtain for the energy density
\begin{eqnarray}
\rho(t)=\rho_{_{0}} \frac{e^{3 \xi_{_{0}} t}}{\left(1+
\frac{H_{_0}}{\xi_{_0}} \,\gamma \left( e^{\, 3 \, \xi_{_{0}}\,
t/2}-1 \right)\right)^{2}},
\end{eqnarray}
where $\rho_{_{0}}=3\,{H}_{_0}^{2}$. As before, for $\gamma < 0$
we have a big rip singularity at a finite value of cosmic time
$t_{_{br}}=\frac{2}{3 \xi_{_{0}}} \, ln(1-\frac{\xi_{_0}}{H_{_0}
\gamma}) > t_{_{0}}=0$.

Note that any additional condition on the system of the field
equations will fix the unknown functions. So for instance, for a
variable $\xi(t)$ we can take the condition $\xi(t)=\xi(\rho(t))$.

Another example in this line is given by the solution obtained by
Barrow~\cite{Barrow 2} for the case $\xi \sim \rho^{1/2}$.
Effectively, Barrow~\cite{Barrow 2} assumed that the viscosity has
a power-law dependence upon the density
\begin{eqnarray}\label{xilaw}
\xi=\alpha \rho^{s}, \,\,\,\,\alpha \geq 0.
\end{eqnarray}
where $\alpha$ and $s$ are constant parameters, and exact
cosmological solutions for a variety of $\xi (\rho)$ in the form
given by equation ~(\ref{xilaw}). In particular, for the case
$s=1/2$, i.e., $\xi =\alpha \rho^{1/2}$, yields a power-law
expansion for the scale factor. Nevertheless, none condition was
imposed upon the parameters $\alpha$ and $\gamma$ in order to
obtain solutions with big rip.

For the case $s=1/2$, the integration of equation~(\ref{Hpunto})
yields the following expression for $H(t)$
\begin{eqnarray}\label{Hdet}
\frac{1}{H}= \frac{1}{H_{0}}-\frac{3}{2}\left( \sqrt{3}\alpha
-\gamma \right)\left( t-t_{0}\right),
\end{eqnarray}
where $H_{0}=H(t=t_{0})$ and $t_{0}$ correspond to the time where
dark component begins to become dominant. The scale factor becomes
\begin{eqnarray}\label{scalefactor}
a(t)=a_{0}\left(1-\frac{t-t_{0}}{t_{br}}\right)^{\frac{2}{3(\gamma
-\sqrt{3}\alpha)}},
\end{eqnarray}
where $a_{0}=a(t=t_{0})$. If we demand to have the occurrence of a
big rip in the future cosmic time then we have the following
constraint on the parameters $\alpha$ and $\gamma$
\begin{eqnarray}\label{constrain}
\sqrt{3}\alpha > \gamma,
\end{eqnarray}
leading the scale factor blow up to infinity at a finite time
$t_{br}>t_{0}$, which expression is
\begin{eqnarray}\label{tbig}
t_{br}=\frac{2}{3(\sqrt{3}\alpha -\gamma)}H_{0}^{-1}.
\end{eqnarray}
In terms of time $t_{br}$, the Hubble parameter is given by
\begin{eqnarray}\label{Hubble}
H(t)= H_{0}\left(1-\frac{t-t_{0}}{t_{br}}\right)^{-1}.
\end{eqnarray}
From the equation~(\ref{tt}) and the parameterized equations
~(\ref{scalefactor}) and ~(\ref{Hubble}) for the scale factor and
Hubble parameter, respectively, we obtain the expression for the
increasing density of the dark component in terms of scale factor
\begin{eqnarray}\label{density}
\rho(a)= 3H_{0}^{2}\left(\frac{a}{a_{0}}\right)^{3(\sqrt{3}\alpha
-\gamma)}.
\end{eqnarray}
We reproduce completely this solution if we put into the
Eq.~(\ref{scalefactor1}) the bulk viscosity given by
\begin{eqnarray}\label{Hubble15}
 \xi(t)= \sqrt{3} \, \alpha
\,H_{0}\left(1-\frac{t-t_{0}}{t_{br}}\right)^{-1}.
\end{eqnarray}

\subsection{The case for $\gamma= 0$}

Notice that the structure of equation~(\ref{Hpunto}) changes if
$\gamma =0$ and $\xi$ is an arbitrary function of the density,
since the quadratic term in $H$ dissapears. Nevertheless, if $\xi
\sim \rho^{1/2}$, the structure of the equation~(\ref{Hpunto}) is
the same for any value of $\gamma$ in the range $0\leq \gamma \leq
2$, except in the case $\sqrt{3}\alpha = \gamma$, where
equation~(\ref{Hpunto}) becomes $\dot{H}=0$. Then, the solution
with $\gamma = 0$ can be obtained directly from the general
solution given by equation ~(\ref{scalefactor}). In this case
there is a big rip singularity at a finite value of cosmic time
$t_{_{br}}=2/( 3 \sqrt{3} H_{_0} \alpha)> t_{_{0}}=0$.

\section{Israel-Stewart-Hiscock theory}
We now consider the dissipative process in the universe within the
framework of the full causal theory of Israel-Stewart-Hiscock. In
this case we have the same Friedmann equations but instead of
equation~(\ref{xi}), we have an equation for the causal evolution
of the bulk viscous pressure, which is given by
\begin{eqnarray}\label{xievolution}
\tau \dot{\Pi}+\Pi= -3\xi H - \frac{1}{2}\tau \Pi
\left(3H+\frac{\dot{\tau}}{\tau}-\frac{\dot{\xi}}{\xi}-\frac{\dot{T}}{T}
\right),
\end{eqnarray}
where $T$ is the temperature and $\tau$ the relaxation time. In
order to close the system we have to give the equation specifying
$T$
\begin{eqnarray}\label{temp}
T= \beta \rho^{r}.
\end{eqnarray}
The the relaxation time is defined by the expression
\begin{eqnarray}\label{tau}
\tau=\frac{\xi}{\rho}=\alpha \rho^{s-1},
\end{eqnarray}
where $\beta \geq 0$. This model imposes the constraint
\begin{eqnarray}\label{r}
r= \frac{\gamma -1}{\gamma} ,
\end{eqnarray}
in order to have the entropy as an state function. Notice that the
above constraint exclude the range $0 < \gamma < 1$, which implies
that quintessence fluids are not allowed in this approach. With
the above assumptions the field equations and the causal evolution
equation for the bulk viscosity lead to the following evolution
equation for $H$~\cite{Maartens}
\begin{widetext}
\begin{eqnarray}\label{Hevolut}
\ddot{H}+ \frac{3}{2}\left( 1+ (1-r)\gamma \right)H\dot{H}
+3^{1-s}\alpha^{-1}H^{2-2s}\dot{H} -\left( 1+ r
\right)H^{-1}\dot{H}^{2}+\frac{9}{4}\left(\gamma-2) \right)H^{3}
+\frac{1}{2}3^{2-s}\alpha^{-1}\gamma H^{4-2s}=0.
\end{eqnarray}
\end{widetext} As in the non causal case we will choose $s=1/2$ and
the above equation becomes
\begin{eqnarray}\label{Hunmedio}
\ddot{H}+ b H\dot{H}-\left( 2- \frac{1}{\gamma} \right)H^{-1}
\dot{H}^{2}+ a H^{3}=0,
\end{eqnarray}
where $a$ is defined by
\begin{eqnarray}\label{a}
a \equiv \frac{9}{4}\left(
\left(1+\frac{2}{\sqrt{3}\alpha}\right)\gamma -2 \right),
\end{eqnarray}
and $b$ by
\begin{eqnarray}\label{b} b \equiv 3 \left( 1+
\frac{1}{\sqrt{3}\alpha} \right).
\end{eqnarray}

Solutions of equation~(\ref{Hunmedio}),were obtained
in~\cite{Mak}. In this work only was considered $\gamma$ in the
range $1\leq \gamma \leq 2$. Some of these solutions presents an
increasing energy density and accelerated expansion.

Inspired in the solution for the Hubble parameter given by
equation~(\ref{Hubble}) in the non causal scheme, we use the
following Ansatz, where for simplicity we take $t_{0}=0$
\begin{eqnarray}\label{Hubble1}
H(t)= A\left(\tau_{br}-t \right)^{-1},
\end{eqnarray}
where $A\equiv H_{0}\tau_{br}$. With this Ansatz the scale factor
$a(t)$ evolutes as
\begin{eqnarray}\label{afactor}
a(t)\sim \left(\tau_{br}-t \right)^{-A},
\end{eqnarray}
and the energy density, $\rho$, of the dark component as a
function of the scale factor becomes
\begin{eqnarray}\label{density}
\rho(a)\sim a^{2/A}.
\end{eqnarray}

%%%%%%%%%%%%%%%%%%%%%%%%%%%%%%%%%%%%%%%%%%%%%%%%%%%%%%%%%%%%%%%%%%%%%%%%%%%%%%
%%%%%%%%%%%%%%%%%%%%%%%%%%%%%%%%%%%%%%%%%%%%%%%%%%%%%%%%%%%%%%%%%%%%%%%%%%%%%%
%%%%%%%%%%%%%%%%%%%%%%%%%%%%%%%%%%%%%%%%%%%%%%%%%%%%%%%%%%%%%%%%%%%%%%%%%%%%%%
%%%%%%%%%%%%%%%%%%%%%%%%%%%%%%%%%%%%%%%%%%%%%%%%%%%%%%%%%%%%%%%%%%%%%%%%%%%%%%
%%%%%%%%%%%%%%%%%%%%%%%%%%%%%%%%%%%%%%%%%%%%%%%%%%%%%%%%%%%%%%%%%%%%%%%%%%%%%%
%%%%%%%%%%%%%%%%%%%%%%%%%%%%%%%%%%%%%%%%%%%%%%%%%%%%%%%%%%%%%%%%%%%%%%%%%%%%%%
%%%%%%%%%%%%%%%%%%%%%%%%%%%%%%%%%%%%%%%%%%%%%%%%%%%%%%%%%%%%%%%%%%%%%%%%%%%%%%
%%%%%%%%%%%%%%%%%%%%%%%%%%%%%%%%%%%%%%%%%%%%%%%%%%%%%%%%%%%%%%%%%%%%%%%%%%%%%%
%%%%%%%%%%%%%%%%%%%%%%%%%%%%%%%%%%%%%%%%%%%%%%%%%%%%%%%%%%%%%%%%%%%%%%%%%%%%%%

Using the Ansatz~(\ref{Hubble1}) in equation~(\ref{Hunmedio}) we
obtain a second grade equation for $A$
\begin{eqnarray}\label{ecforA}
aA^{2}+b A + \frac{1}{\gamma}=0.
\end{eqnarray}
The solutions for $A$ are given by
\begin{eqnarray}\label{Asol}
2A_{\pm}=-\frac{b}{a}\pm \sqrt{\Delta},
\end{eqnarray}
where the discriminant $\Delta$ has the expression:
\begin{eqnarray}\label{Discrim}
\Delta \equiv\left(\frac{b}{a}\right)^{2}-4\left(\frac{1}{a
\gamma}\right).
\end{eqnarray}

Since we are interested only in positive solutions for $A$, the
coefficient $a$ must be negative. We have two cases of interest.

{\bf Case $a<0; \gamma >0$}. In this case only $A_{+}$ correspond
to a solution with big rip. The parameters $\alpha$ and $\gamma$
satisfy the following constraint
\begin{eqnarray}\label{constrain1}
\sqrt{3}\alpha > \gamma \left( 1 - \frac{\gamma}{2} \right)^{-1}.
\end{eqnarray}
Notice that there is no big rip solution if the cosmic fluid
representing the dark component is stiff matter ($\gamma =2$). The
factor $\left( 1 - \frac{\gamma}{2} \right)^{-1}$ is the
correction introduced by the causal thermodynamics to the
constraint given by equation ~(\ref{constrain}). The solution for
$A_{+}$ is given by
\begin{eqnarray}\label{soluciondeA}
A_{+}= \frac{1}{3} \frac{\left(1+\frac{1}{\sqrt{3}\alpha}\right)+
 \left(\frac{1}{3\alpha ^{2}}+
 \frac{2}{\gamma}\right)^{1/2}}{1-\left(1+\frac{2}{\sqrt{3}\alpha}\right)\frac{\gamma}{2}},
\end{eqnarray}
which implies that a big rip will occurs at a time
\begin{eqnarray}\label{soluciondeA}
\tau_{br}= A_{+}H_{0}^{-1}.
\end{eqnarray}
The expressions for $a=a(t)$ and $\rho =\rho (a)$ can be easily
evaluate from equations~(\ref{afactor}) and~(\ref{density}),
respectively.

{\bf Case $a<0; \gamma <0$ }. Since we need $\Delta \geq 0$ in
order to have real solutions, the parameters $\alpha$ and $\gamma$
must satisfy the following constraint
\begin{eqnarray}\label{alfa1}
\sqrt{3}\alpha \leq \sqrt{\frac{\mid \gamma \mid}{2}}.
\end{eqnarray}

If $\Delta =0$, i.e., $\sqrt{3}\alpha = \sqrt{\mid \gamma \mid
/2}$, the solution for $A$, which we shall call $A_{0}$, has the
following expression
\begin{eqnarray}\label{solAo}
A_{0}= \frac{2}{3}
\frac{\left(1+\frac{1}{\sqrt{3}\alpha}\right)}{2+\left(1+\frac{2}{\sqrt{3}\alpha}\right)\mid
\gamma \mid }.
\end{eqnarray}

If $\Delta >0$, the solutions for $A$ can be written as
\begin{eqnarray}\label{solAmas}
A_{\pm}= A_{0} \pm \frac{1}{2}\sqrt{\Delta}.
\end{eqnarray}

\section{Discussion}

Within the framework of the non causal thermodynamics we have
showed that the power law solution, found by Barrow
in~\cite{Barrow 2} for dissipative universes with $\xi =\alpha
\rho^{1/2}$, yields cosmologies which present big rip singularity
when the constraint given in equation~(\ref{constrain}) holds. If
we consider that the dark component is quintessence, i.e., $0\leq
\gamma \leq 2/3$, with a sufficiently large bulk viscosity will
make this quintessence behaves like a phantom energy. In the range
$2/3 > \gamma \geq 2$ it is possible, at least from the
mathematical point of view, to obtain solutions with big rip even
with a matter fluid. It is not clear for us how can be interpreted
a radiation fluid, for example, with a large bulk viscosity
leading to high negative pressures and increasing densities.

At the boundary between the quintessence sector and the phantom
sector, i.e. $\gamma=0$ or $p=-\rho$, also there exist cosmologies
with a big rip singularity.

Using a more accurate approach like the the full causal theory of
Israel-Stewart-Hiscock, we have also found cosmological solutions
with big rip.

If $1\leq \gamma \leq 2$ the parameters $\alpha$ and $\gamma$
satisfy the constraint given in equation~(\ref{constrain1}). Due
to the constraint stiff matter is not allowed. As we mentioned
above this correspond to matter fluids that can lead to a phantom
behavior. Quintessence region are not allowed. If $\gamma <0$ the
cosmological solutions can be computed directly from equations
~(\ref{solAo}) and~(\ref{solAmas}). The main conclusion, in the
context of the full causal thermodynamics, is that in order to
obtain physically reasonable big rip solutions, the dark component
must be phantom energy.

Note added: While this manuscript was being written we noticed
about the work of Brevik and Gorbunova~\cite{Brevik}. The authors
also consider the possibility of big rip in viscous fluids with
$p=w \rho$, by a different formalism. They consider the case where
the bulk viscosity is proportional to the scalar expansion. This
is equivalent to the Barrow's choice $\xi(t) \propto \sqrt{\rho}$,
and then they also reobtained the Barrow's solution, for $\gamma
\neq 0$, considered here. In the framework of the standard Eckart
theory~cite{Eckart}, the authors show that fluids which lie in the
quintessence region ($w>-1$) can reduce its thermodynamical
pressure and cross the barrier $w=-1$, and behave like a phantom
fluid ($w<-1$) with the inclusion of a sufficiently large bulk
viscosity. The case for $\gamma=0$ was not considered by these
authors.

\section{acknowledgements}
NC and SL acknowledge the hospitality of the Physics Department of
Universidad de Concepci\'on where an important part of this work
was done last January. SL acknowledges the hospitality of the
Physics Department of Universidad de Santiago de Chile. We thanks
the suggestion of a new reference given by the referee, in order
to improve the presentation of this paper. We acknowledge the
partial support to this research by CONICYT through grants N$^0$
1051086, 1030469 and 1040624 (MC); N$^0$ 1040229 (NC and SL);
Grant MECESUP USA0108 (NC). It also was supported by the Direccion
de Investigaci\'on de la Universidad del B\'\i o--B\'\i o (MC),
PUCV Grant 123.771/04 (SL).

\end{document}